\documentclass[preprint,final,5p,times,twocolumn]{elsarticle}

\usepackage{rotating,color,subfigure}
\usepackage{alphalph}
\usepackage[T1]{fontenc}
\journal{Physics Letters B}

\usepackage{notoccite}
\usepackage{appendix}
\usepackage{physics}
\usepackage{mathrsfs}
\usepackage{amsmath}
\usepackage{amssymb}
\usepackage[utf8]{inputenc}
\usepackage{xcolor}
\usepackage{graphicx}
\usepackage{dcolumn}
\usepackage{bm}
\usepackage{mathtools,slashed}

\begin{document}


\title{Dark matter: red or blue?}

\author[UoY]{A. Acar}
\author[UoY]{C.~Isaacson}
\author[UoY]{M.~Bashkanov}\ead{mikhail.bashkanov@york.ac.uk}
\author[UoY]{D.P. Watts}
\address[UoY]{School of Physics, Engineering and Technology, Department of Physics, University of York, Heslington, York, Y010 5DD, UK}

\date{\today}

\begin{abstract}
We report the first calculation of light scattering on heavy dark matter (DM) particles. We show that despite the fact that DM has no direct coupling to photons, the light-DM($\gamma\chi$) ($m_\chi \sim 1$ TeV)  cross-section is non-vanishing, albeit small. The cross section, calculated within the Standard Model (SM) framework, is particularly large in the case of heavy Weakly Interacting Massive Particles (WIMP). Combined with astrophysical observation, these results can constrain existing WIMP DM models in favor of lighter DM, $M_\chi<<M_{\mathrm{Planck}}$, (axions, composite DM, etc..) or non-weakly interacting pure gravitational DM. We also show that the energy dependence of light scattering on dark matter should make the DM colored - red in the case of weak-DM and blue for the gravitational-DM, when a white background light is passing through. Gravitational scattering of light on DM particles also leads to non-trivial polarization effects, which might be easier to detect than the deflection of light from the scattering on DM particles, $\gamma\chi\rightarrow\gamma\chi$.
\end{abstract}

\maketitle


\section{\label{sec:Intro} Introduction}
The concept of dark matter (DM) has a 100-year history. There are plenty of astrophysical observations, indicating that ordinary luminous matter covers only 5\% of the energy balance of our Universe and that we should have 4 times as much matter, which we cannot observe directly, {\it matiere obscure}. DM best reveals itself in rotational curves of spiral galaxies, where Newtonian motions of stars far outside the galactic center give us information about the gravitational mass of the galaxy which can be compared to the mass derived from the luminous matter. The disrepancy between the masses derived gravitationally and optically has been attributed to  DM, which produces gravity, but cannot be seen in telescopes. One of the earliest models of the DM, the so-called MACHO's (Massive Compact Halo Objects), assumed large populations of heavy jupiter scale planet systems, forming galactic halos \cite{MACHOS}. In recent years, most variations of such models were discarded \cite{DiscardingMACHOS}. The most prominent DM candidate is currently the WIMP (weakly interacting massive particle). The properties of this hypothetical particle are that it does not interact strongly or electromagnetically, but only weakly and gravitationally \cite{WIMP1}. Current constraints from underground DM search experiments limit the spin-independent DM-nucleon scattering cross section to be ($\sigma/M_{\chi}<3.7  \times 10^{-46}cm^2 \times {M_{\chi}}/{1 \\ \ \mathrm{TeV}}$, for $M_{\chi}>1$ TeV)~\cite{XENON:2025vwd}, where $M_{\chi}$ is the WIMP mass. Very heavy DM candidates, e.g MACHO's, are not restricted in their interactions, but as the DM mass gets lighter there are stronger constrains on possible interaction probabilities. For DM masses of $M_{\chi}\sim 1-10^{14}$~GeV, that usually implies that DM should not interact electromagnetically.

In many DM models it is assumed that the absence of a direct DM-photon vertex implies the inability to probe DM with photons. At some level it is known that this assumption is incorrect. There are several examples, when objects which are forbidden to interact directly with photons, can still interact indirectly due to various loops: Casimir effect \cite{Casimir}, photon interaction with a standard model Dirac neutrino through the loop-induced magnetic moment \cite{NeutrinoPhoton}, even Higgs boson decay into two photons \cite{Marciano}. The latter case is particularly interesting:
the $\Gamma(H\rightarrow \gamma\gamma)=9.2$~keV width is an order of magnitude larger than the $\Gamma(H\rightarrow \mu^+\mu^-)=0.91$~keV, due to Higgs-photon interactions via the $W^{\pm}$ and $t-$quark loops, see Fig.~\ref{Diagrams}. 

In this paper we obtain a first calculation  of how strong the interaction of photons could be with the DM particles and through comparison of the predicted effects with astronomical data, obtain constraints on possible DM candidates. We have tried to minimise the underlying assumptions in our methods and consider two cases: i) when a DM particle can interact weakly and all of its dynamics are defined by conventional standard model physics and ii) when a DM particle can interact only gravitationally. The paper is organised as follows: in Section II we introduce possible weak light-DM scattering processes and calculate their probabilities; in Section III we introduce gravitational light-DM scattering; in Section IV we show results and confront them with existing observational constrains.

\section{Weak Interactions}
\begin{figure}[!h]
\begin{center}
\includegraphics[width=0.47\textwidth,angle=0]{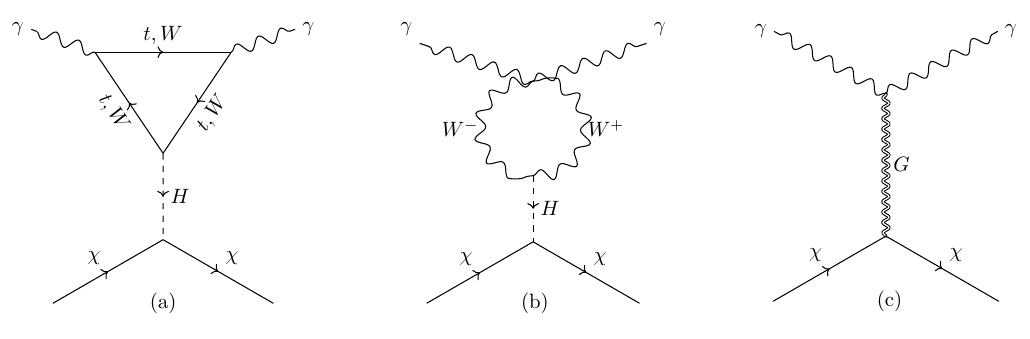}
\end{center}
\caption{All possible diagrams for dark matter-photon scattering propagated by the Higgs boson and Graviton in the unitary gauge.}
\label{Diagrams}
\end{figure}

Since the introduction of weak interaction theory by Weinberg and Salam \cite{Weinberg} people hypothesised about possible extensions beyond the Standard Model.
In recent years there were plenty of other experiments aiming to discover physics beyond the standard model~\cite{FASER}, but no SM violations were discovered so far. Guided by these findings, we assume that any heavy weakly interacting DM particle should interact within the framework of the current Standard Model, without introduction of any additional forces, fields or conservation laws. In this paradigm one can still introduce a WIMP, {\it e.g.} in the form of a very heavy 4th generation neutrino \cite{DarkNeutrino}, a right-handed neutrino \cite{righthanded} or, indeed, any stable DM fermion. In any of these cases a fermionic DM particle would necessarily get its mass through the Yukawa coupling with the Higgs boson. In this scenario, DM would unavoidably interact with light, since DM must interact with the Higgs boson, and the Higgs can decay into two photons. On a diagrammatic level, this process is expressed in Fig.~\ref{Diagrams} (a) and (b), in the unitary gauge~\footnote{The calculations in this work are performed in the unitary gauge  to avoid ghost and Goldstone boson loops.}.  

Following the approach of $H\rightarrow\gamma\gamma$ calculations from Ref.\cite{Marciano}, the DM-photon scattering matrix element can be expressed as:

\begin{align}
i\mathcal{M} = i\mathcal{M_F} + i\mathcal{M_{W}}
\end{align}

\noindent where $i\mathcal{M_F}$ and $i\mathcal{M_{W}}$ are the matrix elements of processes involving the fermion and W-loop respectively. Only the top quark is considered for the fermion loop, as the other particles are too light to contribute to the cross section significantly. The full matrix element integrated over Feynman parameters after dimensional regularization is given by: 


\begin{align}
\nonumber i\mathcal{M} = \frac{\alpha g_w}{4\pi m_w} ((k_1 \cdot k_2)g^{\mu\nu} - k_1^\nu k_2^\mu) \epsilon^*_\mu (k_1) \epsilon_\nu (k_2) \\
\cdot \frac{1}{t-m_H^2+im_H\Gamma_H} [\bar{u}(p_2)\frac{g_w}{2m_w}m_{\chi} u(p_1)] [N_c Q_f^2 I_F + I_W]
\end{align}

\noindent where $\alpha$ is the fine structure constant, $g_w$ is the weak W-boson coupling constant and $m_w$ is the mass of the W-boson. $k_1$, $\epsilon_\mu$ and $k_2$, $\epsilon_\nu$ are the 4-vectors and polarization vectors of outgoing and incoming photons respectively. $m_H$ and  $\Gamma_H$ are the mass and the width of Higgs, $N_c$ is the number of quark colours, $Q_f$ is the charge of the quark depending on its flavour and $m_{\chi}$ is the mass of the dark matter particle. $t$ has the standard definition of the Mandelstam variable, and the $u$ stands for a Dirac spinor. The terms $I_F$ and $I_W$, which come from Feynman parameter integrals, are given by:

\begin{align}
I_F(\beta)= -2\beta(1+ (1-\beta)f(\beta))
\\
I_W(\beta)= 2 + 3\beta+3(2\beta-\beta^2)f(\beta)
\end{align}

where $f(\beta)$ is defined as:

\begin{align}
f(\beta)=
\begin{cases}
  \arcsin^2(\beta^{-1/2}), & \text{if } \beta \ge 1,\\[1ex]
  -\frac{1}{4} [\ln\frac{1+ \sqrt{1-\beta}}{1-\sqrt{1-\beta}} -i\pi]^2,  & \text{if } \beta < 1.
\end{cases}
\end{align}

$\beta$ is $\beta=-4m_w^2/t$ for the W-boson loop, and $\beta=-4m_t^2/t$ for the top quark loop.

\noindent From the square of the full matrix element~\footnote{Since the Higgs is a scalar particle, there is no polarization information transferred: we can expect the $H \chi \chi$ vertex to decouple from the $H\gamma\gamma$ vertex, which makes it easier to obtain the square of the matrix element.}, one can obtain the differential cross section by multiplying it by the relevant flux and phase-space terms\footnote{Full derivation of scattering cross-section calculations can be found in supplemental materials.}:

\begin{align}
\frac{d \sigma}{d \Omega} = \frac{\alpha^2 g_w^4}{(4\pi)^2 m_w^4} \frac{3t^2}{8} \frac{m_\chi^2 (2m_\chi^2 - \frac{t}{2})}{(t-m_H^2)^2 + m_H^2 \Gamma_H^2} \frac {\abs{I_w + N_c Q_f^2 I_f}^2} {64\pi^2 s}
\end{align}

\noindent where {\it s} is the Mandelstam variable.
\begin{figure}[!h]
\begin{center}
\includegraphics[width=0.23\textwidth,angle=0]{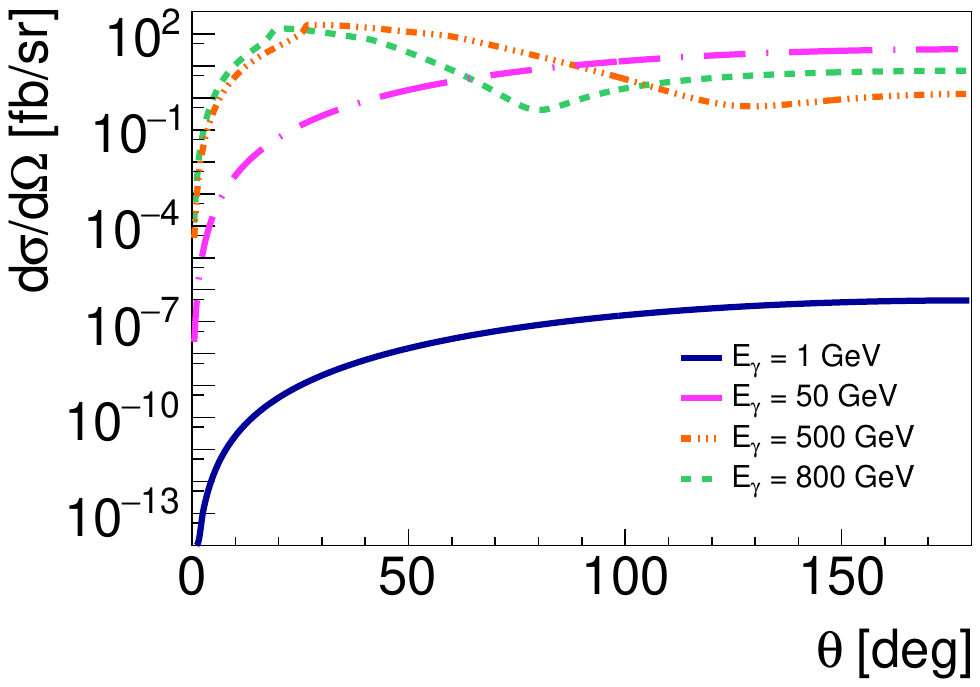}
\includegraphics[width=0.247\textwidth,angle=0]{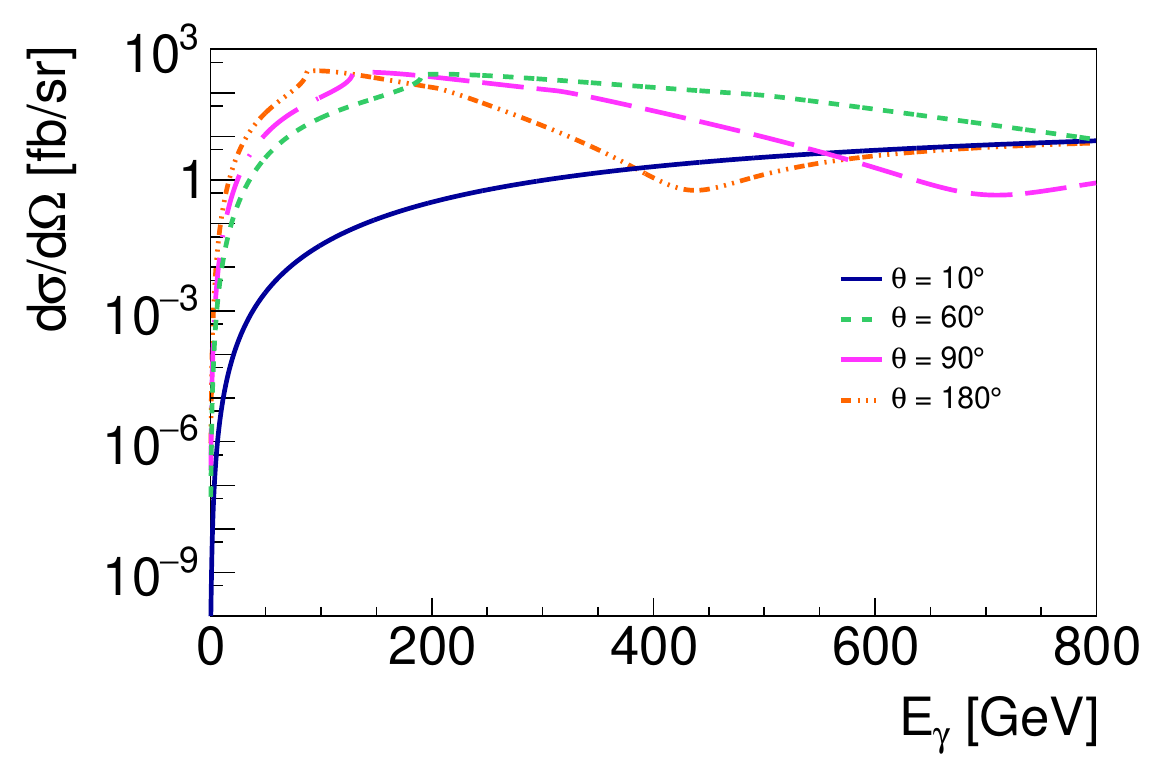}
\end{center}
\caption{Angular(left) and Energy(right) dependence of the differential cross section for photons scattering on $M_{\chi}=1$~TeV WIMP dark matter particles for fixed photon energies of 1~GeV(blue solid), 50~GeV(violet long-dash), 500~GeV(red dash-dot-dot) and 800~GeV(pink short-dash) and fixed photon angles 10 deg (blue solid), 60 deg (red dash-dot-dot), 90 deg(violate long-dash) and 180 deg (pink short dash).}
\label{weak_Edep}
\end{figure}

There are a few things worth noting at this stage.

\begin{figure}[!h]
\begin{center}
\includegraphics[width=0.35\textwidth,angle=0]{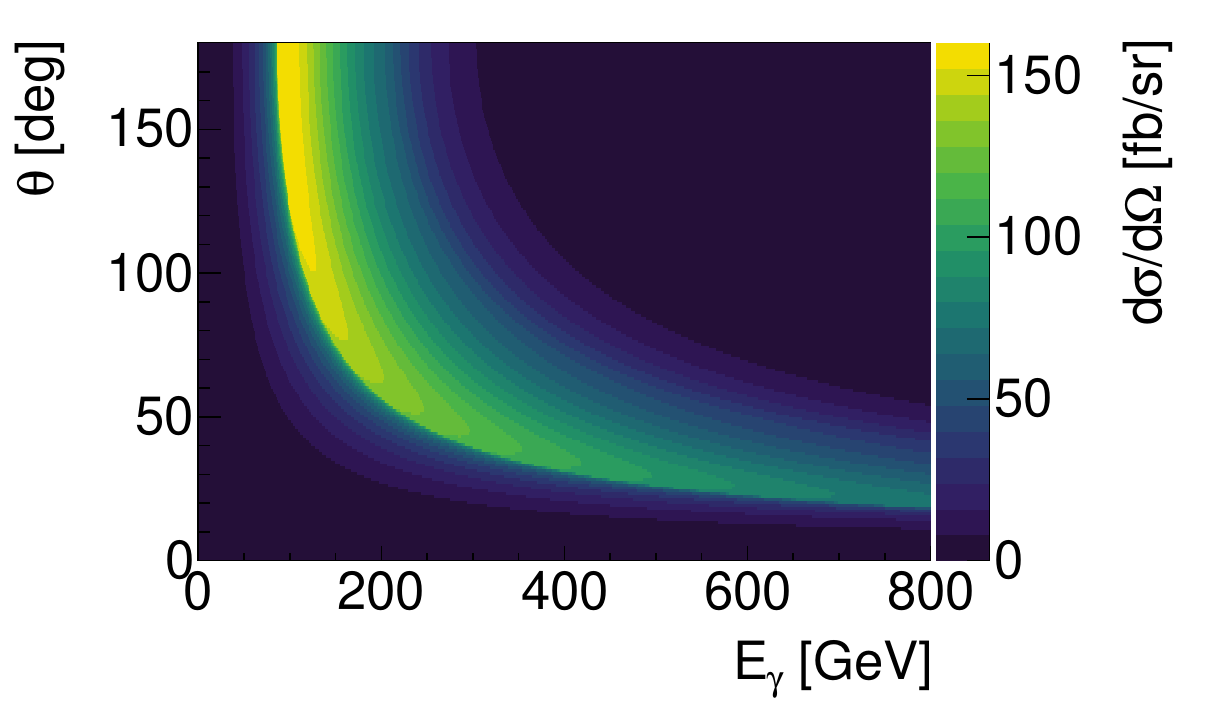}
\end{center}
\caption{ Differential cross-sections for photons scattered on 1~TeV WIMP dark matter particles as a function of photon energy(x-axis) and scattering photon angle (y-axis).}
\label{2Ddiff}
\end{figure}

$\bullet$~The cross section for $\gamma\chi$ scattering rises approximately as $\sigma(\gamma\chi)\sim E_{\gamma}^2$ with the photon energy (for the lower energy photons) reaching a maximum when the loop particles are on their mass shell, Fig.~\ref{2Ddiff}. 

$\bullet$~The propagators (and therefore cross section) for low energy photons are larger for scattering in more backward angular regions, Fig.~\ref{2Ddiff}, meaning white light will become reddish while passing through DM due to the stronger filtering out of higher energy violet photons.

$\bullet$~The scattering cross section is proportional to the mass of DM particle squared, $\sigma(\gamma\chi)\sim M_{\chi}^2$, where the $\chi$ number density scales as $1/M_{\chi}$. The rate of $\gamma\chi$ interactions therefore scales linearly with the DM mass, imposing "non-observation" upper limits on WIMPZilla \cite{Massive} type particles. 

\begin{figure}[!h]
\begin{center}
\includegraphics[width=0.35\textwidth,angle=0]{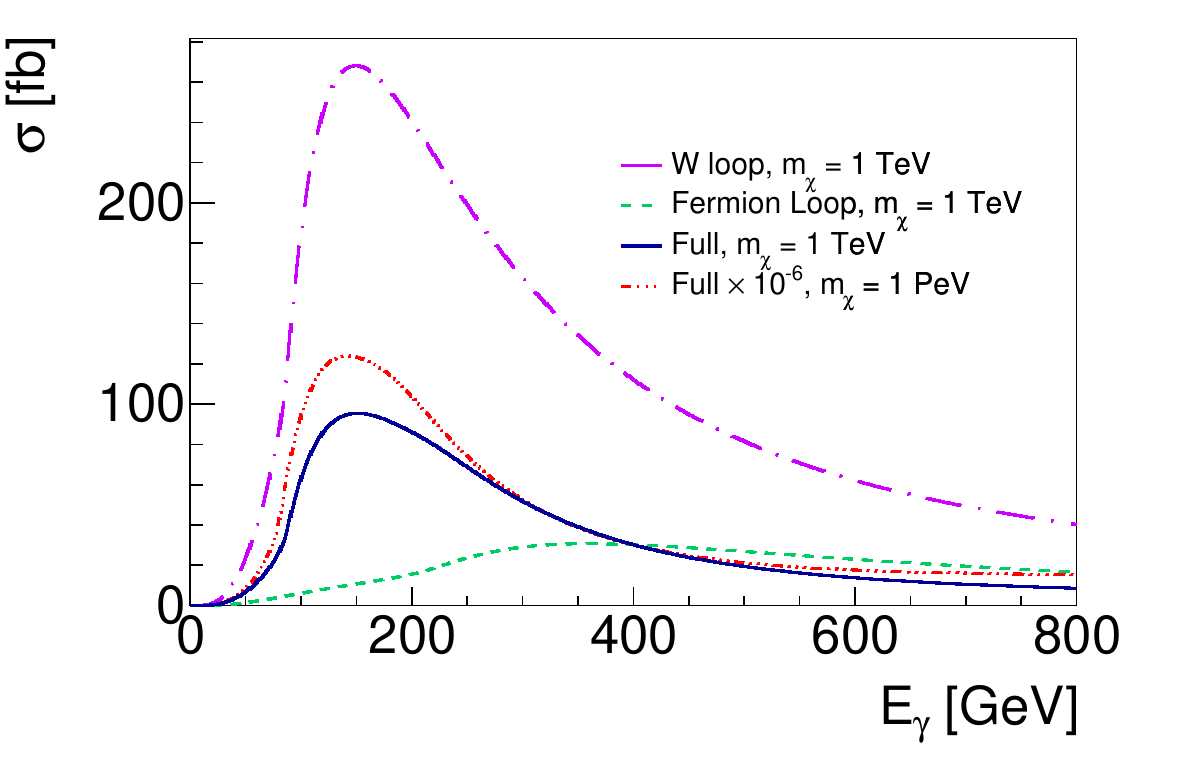}
\end{center}
\caption{The total $\gamma\chi$ scattering cross section alongside with the individual contributions from fermion and W loops for the $M_\chi=1$~TeV.}
\label{Tot_weak}
\end{figure}

$\bullet$~A negative interference between the $W$ and the $t-$quark terms, Fig.~\ref{Tot_weak}, leads to non-trivial energy/angular dependence patterns, seen in Fig~\ref{weak_Edep}, most notably as a dip for back-scattered photons at around $E_\gamma\sim 450$~GeV.

$\bullet$~For a rather standard $M_{\chi}\sim 1$~TeV, the $\sigma(\gamma\chi)\sim 100~fb$ cross section is reasonably large, Fig.~\ref{Tot_weak}, compare to a typical weak cross-sections, e.g. $\sigma(\nu p)\sim 10fb$ at $T_\nu\sim 1~ \mathrm{TeV}$ \cite{icecube}.


\section{Gravitational Interactions}
We have been using gravitational lensing to observe DM clumps from gravity-photon interaction, but much less is known for the single $\gamma\chi$ scattering case. There is a large class of DM models, where DM can interact only gravitationally.
But even in this ultimate case one cannot avoid complete  photo-blindness from DM particles. In this letter we have tried to evaluate the $\gamma\chi$ interaction with a simple perturbative treatment. As expected, we have found a non-zero probability for light to scatter from the DM particles, even for particles which only interact gravitationally. 

Perturbative Quantum Gravity (PQG) is a formalism which treats the force of gravity as an Effective Field Theory. In the low-energy, weak field limit, PQG is generally considered a good approximation for a quantum theory of gravity, see Ref.~\cite{donoghue2012effective} for a review. Instead of working with the highly non-linear field equations of GR, we approximate gravity as a \textit{linear} theory. Linearized general relativity does not pose the same renormalization issues as traditional GR, and thus PQG provides a framework to model quantum effects of gravity, in the low energy limit. In GR (see Ref.~\cite{wald1984general}), space-time is described using a curved metric $g_{\mu\nu}$. The extent of the curvature at a point on the metric corresponds to the strength of the gravitational field at that location. Whereas GR allows for extremely large integrated curvature (eg. near a singularity), PGQ only allows for small perturbations from the flat space-time metric. More precisely, we assume that that $g_{\mu\nu}=\eta_{\mu\nu}+\kappa h_{\mu\nu}$, where $\eta_{\mu\nu}$ is the flat Minkowski metric, $\kappa$ is a constant, and $h_{\mu\nu}$ is the perturbation from the flat metric, $\eta^{\mu \nu}$. This linear perturbation represents the effect of gravity as seen by the metric. 

In our calculations we follow the prescription from Ref.~\cite{Grav} and use the interaction Lagrangian of the form

\begin{align}
    &\mathscr{L}_{int}=-\frac{\kappa}{4} \eta_{\alpha\beta}h^{\alpha\beta}\partial_\mu A_\nu\partial^\mu A^\nu+\frac{\kappa}{4} \eta_{\alpha\beta}h^{\alpha\beta}\partial_\mu A_\nu\partial^\nu A^\mu \nonumber
    \\
    &-\frac{\kappa}{2} h^{\mu\nu}\partial_\mu A_\rho\partial^\rho A_\nu
    +\frac{\kappa}{2} h^{\mu\nu}\partial_\rho A_\mu\partial^\rho A_\nu
    +\frac{\kappa}{2}h^{\mu\nu}\partial_\mu A_\rho\partial_\nu A^\rho \nonumber
    \\
    &-\frac{\kappa}{2} h^{\mu\nu}\partial_\rho A_\mu\partial_\nu A^\rho 
\end{align}

Where $\kappa$ is a constant, and $A^\mu$ is the electromagnetic four-potential vector.
\noindent Within this framework it is then only a matter of applying Feynman Rules.\cite{weinberg1964feynman1,weinberg1964feynman2} to calculate each $\gamma\gamma G$ 
vertex  \cite{prinz2021gravity,bern2002perturbative}. This yields 10 terms as follows:

\begin{align}
    &V_{k_1,k_2}=\frac{i\kappa}{2}[k_1\cdot k_2(\eta^{\rho\alpha}\eta^{\beta\sigma}-\eta^{\rho\beta}\eta^{\alpha\sigma}-\eta^{\rho\sigma}\eta^{\alpha\beta}) 
    +\eta^{\rho\sigma}k_1^\beta 
    k_2^\alpha \nonumber
    \\
    &+\eta^{\alpha\beta}k_1^\rho k_2^\sigma-\eta^{\alpha\beta}k_1^\sigma k_2^\rho 
    -(\eta^{\alpha\rho}k_2^\sigma k_1^\beta+\eta^{\beta\rho}k_2^\rho k_1^\beta+\eta^{\beta\rho}k_1^\sigma k_2^\alpha \nonumber
    \\
    &-\eta^{\beta\sigma}k_1^\rho k_2^\alpha)]
\end{align}


\noindent where $k$ denotes relativistic momentum, and the subscripts 1 and 2 correspond to the incoming and outgoing particles respectively, similar to Ref.~ \cite{ratzel2016effect}

We consider the diagram on Fig~\ref{Diagrams}(c) for the photon scattering on a dark matter particle, assuming the simplest possible case when the DM is scalar, which is true for many gravitational-DM models.

We choose basis vectors for the photon polarisation~\cite{peet1970cross},  $e^\parallel$ with the plane spanned by incoming and outgoing photons, and the other vector, $e^\perp$, as orthogonal to this scattering plane. Following Ref. \cite{boccaletti1969photon}, the matrix element can be written as:

\begin{align}
    &\bra{K_2,P_2}M\ket{K_1,P_1}=\frac{\lambda^2\delta^4(K_2+P_2-K_1-P_1)}{4(2\pi)^2\sqrt{K_1^0K_2^0P_1^0P_2^0}}\nonumber \\
    &\frac{4}{-2K_1K_2}\cdot
    \{(e_{(\rho}^2K_\alpha^2-e_\alpha^2 K_\rho^2)(e_\sigma^1K^\alpha_1-e_1^\alpha K_{\sigma)}^1)\nonumber\\
    &-\frac{1}{4}\delta_{\rho\sigma}(e^2_\alpha K^2_\beta-e^2_\beta K^2_\alpha)(e^\alpha_1K^\beta_1-e^\beta_1K^\alpha_1)\}\nonumber\\
    &\cdot\left[p^{(\rho}_1p^{\sigma)}_2-\frac{1}{2}m^2\delta^{\rho\sigma}\right]
\end{align}


\noindent where Greek indices represent components of 4-vectors and the numbers $1,2$ correspond to initial and final particles respectively,  $K_1,K_2$ denote the 4-momentum of the initial and final photon respectively, and $P_1,P_2$ denote that of the initial and final scalar particles in the center-of-mass (CMS) frame. The quantity $\lambda=\sqrt{8\pi G}$ governs the strength of the gravitational interaction and $\delta^{\rho\sigma}$ is the usual Kronecker delta and $A_{(\rho}B_{\sigma)}$ denotes symmetrization with respect to the indices $\rho$ and $\sigma$.

Eq. (9) above leads us to the polarization-dependent differential cross sections\footnote{Full derivation of scattering cross-section calculations can be found in supplemental materials.}, see Fig. \ref{Diff_grav}

\begin{equation}
    \begin{split}
        \frac{d\sigma_{\parallel \parallel}}{d\Omega}=\frac{\lambda^4(K+P)^4 \text{cos}^2\theta}{16(2\pi)^2P^2(1-\text{cos}\theta)^2}\\
        \frac{d\sigma_{\perp\perp}}{d\Omega}=\frac{\lambda^4(K+P)^4}{16(2\pi)^2P^2(1-\text{cos}\theta)^2}
    \end{split}
\end{equation}

Where $K$ and $P$ are the energies of the photon and dark matter respectively. It is easy to see that the coefficient of the differential cross section is proportional to $G^2$ and (assuming that the mass of the DM particle is much larger than the photon energy). In our weak field approximation, we should consider only small $P$, and thus, we expect very small cross sections in reasonable scattering scenarios.

\begin{figure}[!h]
\begin{center}
\includegraphics[width=0.35\textwidth,angle=0]{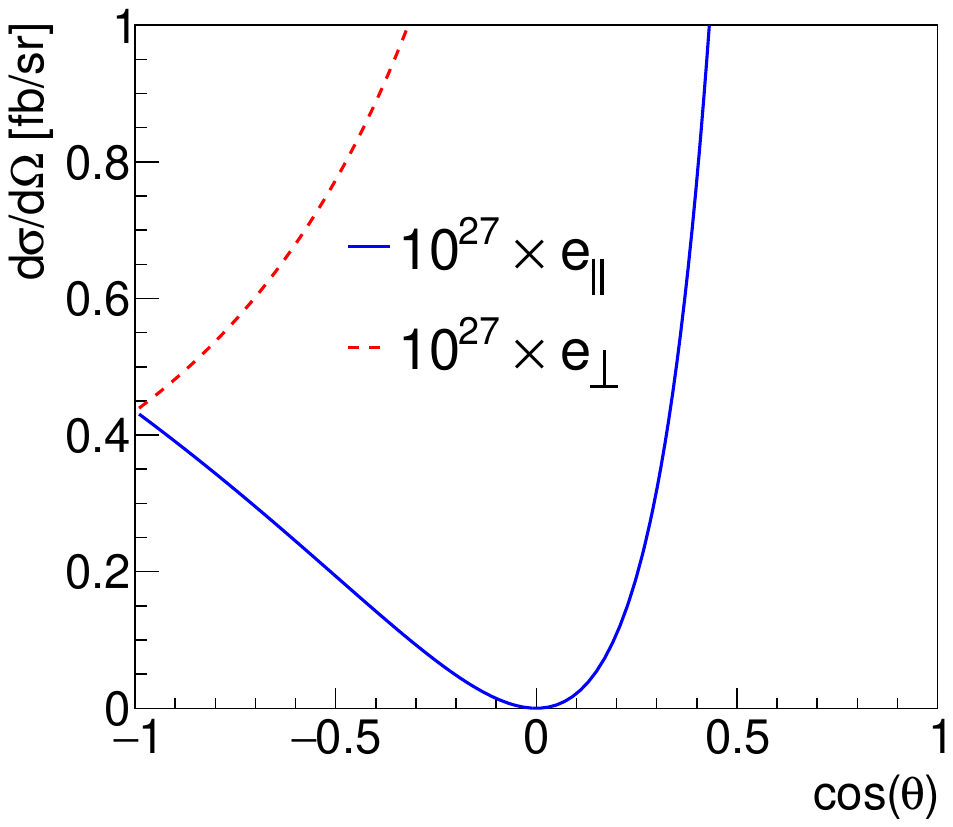}
\end{center}
\caption{The differential $\gamma\chi$ scattering cross section via gravitational interaction for the $M_\chi=10^{19}$~GeV as a function of photon scattering angle for two photon polarisations, using $E_{\gamma} = 200$ GeV. (Note the $10^{27}$ scale)}
\label{Diff_grav}
\end{figure}

There are a few things worth noting about gravitational results as well.

$\bullet$~The cross section for  $\gamma\chi$ scattering rises approximately as $\sigma(\gamma\chi)\sim E_{\gamma}^2$ with the photon energy,
similar to WIMP case, since we only consider small P.

$\bullet$~The cross-section is continuously rising and shows little structure, in contrast to the WIMP case. 

$\bullet$~To maximise the propagator, low energy photons have to scatter forwards, Fig.~\ref{Diff_grav}. Combined with the $\sim E_{\gamma}^2$ energy dependence gives expectation of a white light source becoming more bluish. This picture is in stark contrast to the WIMP case and indicates that light scattering by DM may offer sensitivities to the nature of the DM particles. 

$\bullet$~The cross-section is roughly proportional to the mass of DM particles squared, $\sigma(\gamma\chi)\sim M_{\chi}^2$. Very similar to the WIMP results. Since some of the objects, like quantum black holes~\cite{Bekenstein:1995ju}, can be really heavy it may impose an upper limit on such DM models: if we are considering very heavy gravitationally interacting DM particles, the non-detection of this process, $\gamma\chi \rightarrow \gamma \chi$, can impose limits on how heavy the particles can be.
The current upper limit is $5M_{\odot}$, \cite{PDGDarkMatter} \cite{PDGDarkMatter2} However, this limit is likely to be weaker than the astrophysical microlensing constrains~\cite{Microlensing}.

$\bullet$~Large angle scattering leads to non-trivial polarization dependence, Fig.~\ref{Diff_grav}. Since the cross-section is small, it is highly unlikely that we will be able to measure flux alteration due to $\gamma\chi$ gravitational scattering in the foreseeable future. However, polarisation effects may offer different (improved by $\sim$ 15 orders of magnitude) measurement sensitivities and systematics compared to flux measurements~\footnote{Polarisation effects depends on amplitude linearly, while flux depends quadratically}.

\section{Results}

\begin{figure}[!h]
\begin{center}
\includegraphics[width=0.5\textwidth,angle=0]{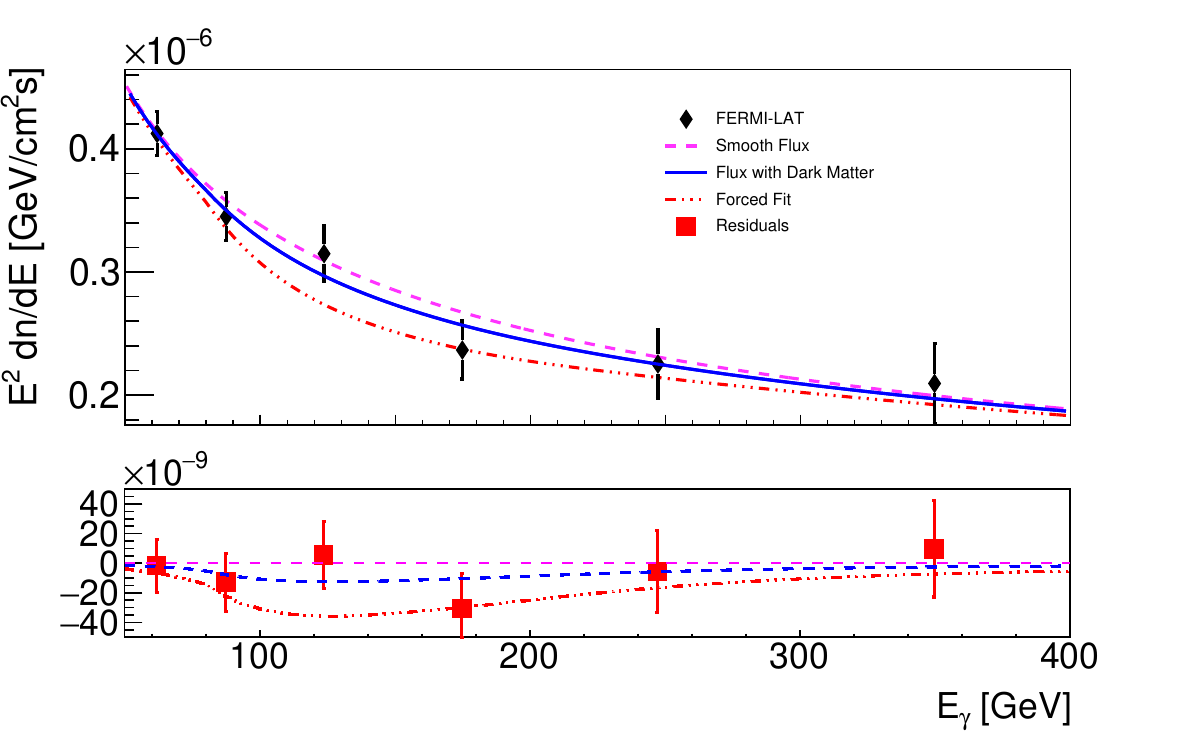}
\end{center}
\caption{\underline{Top} Photon flux from the Galactic centre measured by Fermi-LAT \cite{FermiLAT} in comparison to a fit with (solid blue) and without(dash pink) hypothesis. A solution which was forced to go through the 175 GeV Fermi-LAT point is shown as red dash-dot-dot curve. \underline{Bottom} Residuals relative to no DM fit curve. the colour coding is the same as for the top graph.}
\label{Flux}
\end{figure}

After calculating the cross sections, one can check if the effect of light scattering on DM is detectable and which constrains on  DM mass it can impose. To evaluate this, we have assumed that we have a smooth photon flux coming from the centre of our galaxy and that a DM density distribution follow the NFW
profile \cite{NFW}. Under such assumptions, the light travelling from the centre of galaxy to the Earth will encounter additional scattering on DM leading to a dip in the measured photon flux spectrum, Fig.~\ref{Flux}. By taking the measured photon flux from Fermi-LAT \cite{FermiLAT} we can perform a fit, constraining the maximum possible dip depth, which is consistent with the data, and extract the upper limit for the $M_{\chi}<5.0\cdot 10^{19}$~GeV. On Fig.~\ref{Flux} we show the extracted smooth photon flux background from the fit including a DM component (pink dashed), compared to a fit where the scattering on a DM is also allowed (blue solid). We have also demonstrated the exaggerated version of a fit where we forced it to go through the $E_{\gamma}=175$~GeV data point to emphasize the effect\footnote{The unaffected photon flux was assumed to be a smooth function described by splines. The shape of the unaffected photon flux was chosen to be the same for both normal and forced fits.}.

While the  upper limit for the $M_{\chi}$ is quite high, $\sim$ Plank mass, it is still within WIMPZilla's mass range~\cite{Massive,Transplanckian}. 

It is worth noting that the upper limit for the $M_{\chi}$ in this assumption is solely dependent on the accuracy of the experimental data for the photon spectrum with the large quoted error at $\sigma(E_{\gamma}=175$ GeV)$=10\%$ for a $\sim 100$~GeV bin being the limiting factor. It is anticipated that data with greatly improved accuracy will become available for future analysis~\cite{CubeSat}.

Stronger upper limits can be achieved if we would consider extragalactical sources of high-energy photons. 
For the most distant galaxies \footnote{12 bn l.y. away}, the moderate $M_{\chi}=10^6$~GeV would lead to a 10\% attenuation of flux due to light interaction with the weak DM~\footnote{Assuming an average DM density in the universe of $1.2\cdot 10^{-6}~GeV/cm^3$}. Deriving constraints from this method, however, would require a precise sky survey with a well benchmarked light source and incorporation of detailed and accurate DM density distributions for the luminosity evaluations of each light source.

\section{\label{sec:final} Summary}
We have performed calculations of light-DM, $\gamma\chi$ scattering under two scenarios, when DM is due to a weakly interacting particle which behaves within a standard model paradigm and when DM is a purely gravitational object. For the low energy photons, the energy dependence of the cross-section is very similar, but the angular dependences have different behaviours. The light tends to scatter more backward in the weak case and more forward in the gravitational case - colouring DM-skies in red or blue respectively. The energy dependences of the cross-sections also show differences for the two scenarios. In addition,  it is shown that photon scattering on DM via graviton exchange may lead to very non-trivial polarisation effects,  making it possible to discriminate these two possibilities with future experiments. The work motivates  efforts to better determine absolute flux measurements for high energy gammas passing through dense dark matter regions in future experiments, which would offer new routes to set upper limits on heavy WIMP (WIMPZilla) dark matter scenarios.










\section{Acknowledgements}

We want to thank Fermi-LAT collaboration for providing us their data. We also want to thank Bernard Kay for fruitful discussions regarding gravitational scattering of the light. This work has been supported by the U.K. STFC (ST/L00478X/1, ST/T002077/1, ST/L005824/1, 57071/1, 50727/1 ) grants. After publication all data will be available at Pure@York~\cite{Pure}.

\bibliographystyle{elsarticle-num}   
\bibliography{references_v3}
\clearpage
\onecolumn
\section{Appendix}
\subsection{Gravitationally and Weakly Interacting Dark Matter}
\begin{figure}[!h]
\begin{center}
\includegraphics[width=0.8\textwidth,angle=0]{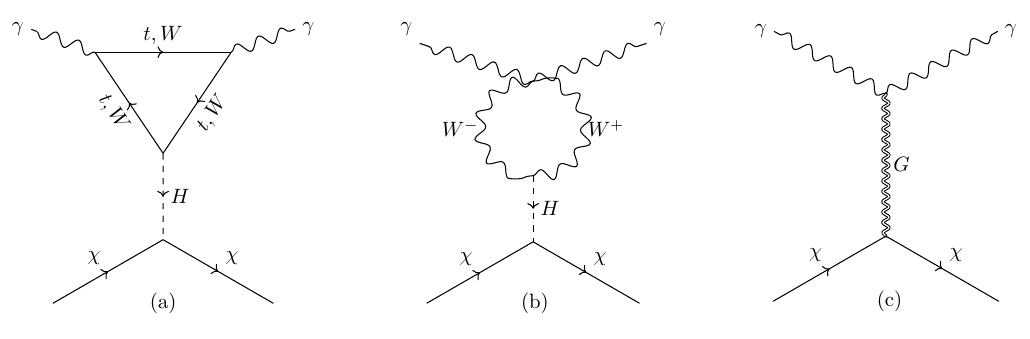}
\end{center}
\caption{Simplest possible diagrams for dark matter-photon scattering propagated by the SM Higgs boson and Graviton in the unitary gauge.}
\end{figure}

We consider  diagrams above to calculate dark matter - photon scattering. We start by going through the weak interactions shown in Figure \ref{Diagrams} (a) and (b), followed by the gravitational interaction shown in (c). Below is a list of common variables that are used throughout the calculations in the following sections and their definitions. 
\\
\\
\\
$k_{1,2}$: 4-momentum of outgoing and incoming photons, respectively.
\\
\\
$p_{1,2}$: 4-momentum of outgoing and incoming dark-matter particles, respectively.
\\
\\
$g_w$:  weak W boson coupling constant.
 \\
 \\
$m_w$:  mass of the W boson.
\\
\\
$m_H$ : mass of the Higgs boson.
 \\
 \\
$\Gamma_H$: width of the Higgs boson.
  \\
  \\
$m_{\chi}$: mass of the dark matter particle.
\\
 
\subsubsection{Weakly Interacting DM}
The full matrix element for the top-quark loop diagram \ref{Diagrams} (a) is given by:

\begin{align}
\label{FME}
    \nonumber i\mathcal{M}_q = \epsilon^{*\mu} (k_1) \epsilon^{\nu} (k_2) N_c Q_f^2 &\int \frac{d^4 l}{(2\pi)^4} \frac{Tr[i(\slashed{l}-\slashed{k_1} + m_f)}{(l-k_1)^2-m_f^2} ie\gamma^\nu \frac{i(\slashed{l}+m_f)}{l^2-m_f^2} \cdot ie\gamma^\mu \frac{i(\slashed{l}+ \slashed{k_2}+ m_f)]}{(l+k_2)^2-m_f^2} \frac{-ig_w m_f}{2m_w} 
    \\
    & \frac{i}{t^2-m_H^2 +im_H\Gamma_H}  \cdot \left[\bar{u}(p_2) \frac{-ig_w}{2m_w} m_\chi u(p_1) \right] 
\end{align}

Where $N_c$ is the number of quark colors, $Q_f$ is the charge of the quark depending on its flavor, $m_f$ is the mass of the top quark in the loop and $\bar{u}$ is the adjoint Dirac spinor. $t$ is the Mandelstam variable. 

Let's look at the trace term in $\mathcal{M}_q$:
\begin{align}
    Tr[i(\slashed{l}-\slashed{k_1} &+ m_f)ie\gamma^\nu i (\slashed{l}+m_f)ie\gamma^\mu(\slashed{l}+\slashed{k_2}+m_f)] \nonumber
    \\
    & =4 ie^2m_f ((4l^\mu l^\nu - g^{\mu\nu} l^2) + 2(l^\mu k_2^\nu - k_1^\mu l^\nu) 
    - k_1^\mu k_2^\nu + k_1^\nu  k_2^\mu + (m_f^2-(k_1 \cdot k_2))g^{\mu \nu})
\end{align}

We perform these trace calculations in spinor space and use $\slashed{k_1}^2 = k_1^2$, with standard gamma matrix identities. 

In order to simplify the products in the denominator, we can utilize Feynman parameters. The Feynman parameterization technique is used to simplify denominators that consist of products of different variables:

\begin{align}
    \frac{1}{CF} &= \int^1_0 \frac{dz}{(zC + (1-z)F)^2}
    \\
    \frac{1}{C_1 ... C_n} &= (n-1)! \int^1_0 dz_1 ... \int^1_0 dz_n \frac{\delta(1- \sum^n_{k=1} z_k)}{(\sum^n_{k=1} z_k C_k)^n}
\end{align}

$C$ and $F$ are of the form $(a+q^2)^{-1}$, where $a$ is a constant \cite{PeskinSchroeder1995}. The Feynman parameters $z_n$, have the property:

\begin{align} \label{FP}
    z_1 + z_2 + ... + z_n = 1
\end{align}
Using this technique, the denominator, $D$, becomes:
\begin{align}
    D &= {((l-k_1)^2-m_f^2)(l^2-m_f^2)((l+k_2)^2-m_f^2))} \nonumber
    \\
      &=z_1 ((l-k_1)^2-m_f^2) + z_2 (l^2 - m_f^2) + z_3 ((l+k_2)^2-m_f^2) \nonumber
    \\
      &=z_1(l^2 - 2l\cdot k_1 + k_1^2 - m_f^2) + z_2l^2 -z_2m_f^2 + z_3(l^2 + 2l\cdot k_2 + k_2^2 - m_f^2) \nonumber
    \\
    &=l^2 + l(-2z_1k_1 + 2z_3k_2) + z_1k_1^2 + z_3k_2^2 - m_f^2
\end{align}

Then we can denote:

\begin{align}
    A &= z_3k_2 -z_1k_1
    \\
    B & = z_1k_1^2 + z_3k_2^2 - m_f^2
\end{align}

Finally, 

\begin{align}
    D &= (l+A)^2 + B - A^2  \nonumber
    \\
      &=(l+z_3k_2 - z_1k_1)^2 - (z_3^2k_2^2-2z_3z_1(k_2\cdot k_1) + z_1^2k^2) + z_1k_1^2 + z_3k_2^2 - m_f^2 \nonumber
    \\
     &=(l+z_3k_2 -z_1k_1)^2 + 2z_3z_1 (k_2 \cdot k_1) - m_f^2
\end{align}

where in the last line we remember that $k_2^2 = k_1^2 = 0$, since these are 4-momenta for photons. We define $\Delta$ as:

\begin{align}
    \Delta = 2z_3z_1(k_2 \cdot k_1) - m_f^2.
\end{align}

We can then make a shift in variable $l$, and define it as $L$:

\begin{align}
     L &= l+ A = l+ z_3k_2 - z_1k_1
    \\
    D &=  L^2 + \Delta
\end{align}

We note that $D$ only depends on the magnitude of $ L$, so we can say:

\begin{align}
    &\int d^4L \frac{L^\mu}{D^3} = 0
    \\
    &\int \frac{d^4L}{D^3} L^\mu L^\nu = \int \frac{d^4L}{D^3} \frac{1}{4} \frac{g^{\mu \nu}L^2}{D^3}
\end{align}

Then we need to put the numerator in a useful form:

\begin{align}
     Tr[i(\slashed{l}-\slashed{k_1} &+ m_f)ie\gamma^\nu i (\slashed{l}+m_f)ie\gamma^\mu(\slashed{l}+\slashed{k_2}+m_f)]  \nonumber
     \\
     &= 4ie^2 m_f (4( L-z_3k_2 +z_1k_1)^\mu ( L-z_3k_2 +z_1k_1)^\nu - g^{\mu \nu}( L-z_3k_2+z_1k_1)^2 \nonumber
    \\
    &+2(( L-z_3k_2+z_1k_1)^\mu k_2^\nu - k_1^\mu ( L-z_3k_2+z_1k_1)^\nu) - k_1^\mu k_2^\nu + k_1^\nu k_2^\mu \nonumber
    \\    
    &+ (m_f^2 - (k_1 \cdot k_2))g^{\mu \nu}) \nonumber
    \\ \nonumber
    \\ \nonumber
    &= 4ie^2 m_f (4  L^\mu  L^\nu + k_2^\mu k_2^\nu(4z_3^2 - 2z_3)+k_1^\mu k_1^\nu (4z_1^2 - 2z_1) - 4z_3z_1k_2^\mu k_1^\nu)+ k_1^\mu k_2^\nu (-4z_1z_3 + 2z_1 + 2z_3) \nonumber
    \\
    & + 2z_1 z_3k_1k_2 g^{\mu \nu} - k_1^\mu k_2^\nu + k_1^\nu k_2^\mu + (mf_2-(k_1 \cdot k_2))g^{\mu \nu} -  L^2g^{\mu \nu})
\end{align}

Our integral in $\mathcal{M}_q$ will concern terms that contain $L$:

\begin{align}
    \int^1_{z_1} \int^{1-z_1}_0 dz_1 dz_3 &\int \frac{d^4L} {(2\pi)^4} ie^2 4m_f 
    \frac{2[4L^\mu L^\nu + k_2^\mu k_2^\nu (4z_3^2 - 2z_3) + k_1^\mu k_1^\nu (4z_1^2 - 2z_1) - 4z_3 z_1 k_2^\mu k_1^\nu]}{(L^2+\Delta)^3} \nonumber
    \\
    &+\frac{2[k_1^\mu k_2^\nu(-4z_1z_3 + 2z_1 + 2z_3) + 2z_1z_3(k_2 \cdot k_1)g^{\mu \nu}-k_1^\mu k_2^\nu + k_1^\nu k_2^\mu]}{(L^2 + \Delta)^3} \nonumber
    \\
    & + \frac{2[(m_f^2-(k_1 \cdot k_2))g^{\mu \nu}-L^2 g^{\mu \nu}]}{(L^2 + \Delta)^3}
\end{align}
We can solve this integral using Wick rotation in d-dimensional space. First, we focus on terms containing $L$ in the numerator:

\begin{align}
    I_1 = \int \frac{d^4L}{(2\pi)^4}\frac{2(\frac{4}{d}-1) g^{\mu \nu} L^2}{(L^2+\Delta)^3}
\end{align}

Where we use:

\begin{align}
    g_{\mu \nu} L^\mu L^\nu = \frac{L^2}{d}g_{\mu \nu}g^{\mu \nu}
\end{align}

Then we apply the Wick rotation:

\begin{align}
    L^0 \rightarrow i L^0_d
\end{align}
The integral becomes:

\begin{align}
    I_1 = i \int dL \frac{2\pi^{d/2}}{\Gamma(d/2)} \frac{2(\frac{4}{d}-1)g^{\mu \nu}(L^2 L^{d-1})}{(2\pi)^d (L^2+ \Delta)^3} 
\end{align}

Note that:

\begin{align}
    \int d^d L = \int d \Omega_d = \int dL \ L^{d-1}
\end{align}

Then, we can further simplify by remembering:

\begin{align}
    \int dL \frac{L^{d+1}}{(L^2 + \Delta)^3} = \frac{\Delta^{d/2 -2}}{2} \frac{\Gamma(2-d/2) \Gamma(1+d/2)}{\Gamma(3)}
\end{align}

Substitute this into the main integral:

\begin{align}
    I_1 = 2i(\frac{4}{d}-1)g^{\mu \nu} \frac{d}{4^{d/2}}\frac{1}{2\pi^{d/2}} \Delta^{d/2-2} \frac{\Gamma(2-d/2)}{\Gamma(3)}
\end{align}

We can say:

\begin{align}
    \Gamma(2-d/2) = \Gamma(\epsilon), \quad 2\epsilon = 4-d
\end{align}

Substituting this into previous expression:

\begin{align}
    I_1 = 2i (\frac{4}{d}-1)g^{\mu \nu } \frac{d}{4^{d/2}} \frac{1}{2\pi^{d/2}} \Delta^{- \epsilon} \frac{\Gamma(\epsilon)}{\Gamma(3)}
\end{align}

We can expand:

\begin{align}
    \left(\frac{4}{d}-1 \right)d = 4 -d 
\end{align}

And we get:

\begin{align}
    I_1 &= \frac{i2g^{\mu \nu} \epsilon}{(4\pi)^{d/2}} \Delta^{-\epsilon} \frac{\Gamma(\epsilon)}{\Gamma(3)} \nonumber
    \\
    &= \frac{i2g^{\mu \nu} \epsilon}{(4\pi)^{2}} \left (\frac{\Delta}{4\pi} \right)^{-\epsilon} \frac{\Gamma(\epsilon)}{\Gamma(3)}
\end{align}

Then we can expand the $\Gamma$-function:

\begin{align}
    \Gamma(x) = \frac{1}{x} - \gamma + O(x)
\end{align}

where $\gamma$ is the Euler-Mascheroni constant.

\begin{align}
    I_1 = \frac{i2g^{\mu \nu} \epsilon}{(4\pi)^{2}} \left (\frac{\Delta}{4\pi} \right)^{-\epsilon} \left( \frac{1}{\epsilon} - \gamma \right)\frac{1}{\Gamma(3)}
\end{align}
We can plug in $\Gamma(3) = 2$:

\begin{align}
     I_1 = \frac{i2g^{\mu \nu} \epsilon}{(4\pi)^{2}} \left (\frac{\Delta}{4\pi} \right)^{-\epsilon} \frac{\left( \frac{1}{\epsilon} - \gamma \right)}{2}
\end{align}

If we evaluate this in the limit $\epsilon \rightarrow 0$, we get:

\begin{align}
    I_1 = \frac{i g^{\mu \nu}}{16 \pi^2}
\end{align}

Then we can evaluate the rest of the integral using the same technique:

\begin{align}
    I_2 = \int^1_{z=1} \int^{1-z_1}_0 &\frac{d^4L}{(2\pi)^4} 8ie^2 m_f [(4z_3^2 - 2z_1)k_1^\nu k_1^\mu + (4z_1^2-2z_3)k_2^\nu k_2^\mu - 4z_3 z_1 k_2^\mu k_1^\nu \nonumber
    \\
    &k_1^\mu k_2^\nu (-4z_1z_3 + 2z_1 + 2z_3) + 2z_1z_3(k_1 \cdot k_2)g^{\mu \nu} - k_1^\mu k_2^\nu + k_1^ \nu k_2^\mu +  \nonumber
    \\
    &(m_f^2 - (k_1 \cdot k_2))g^{\mu \nu}] \frac{1}{(L^2+\Delta)^3}
\end{align}
Let's perform the Wick rotation and calculate the term in the integral where $L$ is present:

\begin{align}
    \int \frac{d^4L}{(2\pi)^4} \frac{1}{(t^2+\Delta)^3} &= i  \frac{1}{(2\pi)^d} \int \frac{(2 \pi)^{d/2}}{\Gamma(d/2)} dL \frac{L^{d-1}}{(L^2+\Delta)^3} \nonumber
    \\
    & = \frac{i \pi^2}{(2\pi)^4} \frac{\Gamma(1)}{\Gamma(3)} \frac{1}{\Delta} \nonumber
    \\
    &= \frac{i}{32\pi^2\Delta}
\end{align}

Re-write the matrix element, and use Ward identities ($k_1^\mu e^\lambda_\mu = 0, k_1=k_2^\nu e^\lambda_\nu = 0$) to simplify terms:
\begin{align}
i \mathcal{M}_q &= 
\epsilon^{*\mu}(k_1)\,\epsilon^\nu(k_2) \,
N_c Q_f^2 
\left( \frac{-i g_w}{2 m_w} \right) 
m_f \, i e^2 
\Bigg[ 
    \frac{4i\, g_{\mu \nu}}{16\pi^2} 
    + \frac{i}{32\pi^2 \Delta} 
      \cdot 
      \bigg(
        \int_{z_1=0}^1 dz_1 
        \int_{0}^{1-z_1} dz_3 
        8 \big(
            (4 z_3^2 - 2 z_1) k_1^\nu k_1^\mu \nonumber
            \\
            &- 4 z_3 z_1 k_2^\mu k_1^\nu
            + (4 z_1^2 - 2 z_3) k_2^\nu k_2^\mu
            + k_1^\mu k_2^\nu (-4 z_1 z_3 + 2 z_1 + 2 z_3)
            + 2 z_1 z_3 (k_1 \cdot k_2) g^{\mu \nu} \nonumber
            \\
            &- k_1^\mu k_2^\nu
            + k_1^\nu k_2^\mu
            + (m_f^2 - (k_1 \cdot k_2)) g^{\mu \nu} \nonumber
        \big)
      \bigg)
\Bigg]
\\
 &= -  \epsilon^{*\mu}(k_1) \epsilon^{\nu} (k_2) N_c Q_f^2 \frac{g_w}{2m_w} m_f e^2 \int^1_{z_1=0} dz_1 \int^{1-z_3}_{0} dz_3 \big(\frac{g_{\mu \nu}}{4 \pi^2} + \frac{1}{4 \pi^2 \Delta} [(1-4z_1z_3) k_1^\nu k_2^ \mu \nonumber
       \\
       &+ g^{\mu \nu} (2z_1z_3 (k_1 \cdot k_2) + m_f^2 - (k_1 \cdot k_2))] \big)
\end{align}

We can then define $\beta$:
\begin{align}
\beta = \frac{2{m_f^2}}{(p \cdot k)} = -\frac{4 m_f^2}{t} 
\end{align}

The integral expression evaluates to:
\begin{align}
     I_f = \int^1_{z_1=0} dz_1 \int^{1-z_3}_{0} dz_3 \frac{g_{\mu \nu}}{4 \pi^2} + &\frac{1}{4 \pi^2 \Delta} [(1-4z_1z_3) k_1^\nu k_2^ \mu + g^{\mu \nu} (2z_1z_3 (k_1 \cdot k_2) + m_f^2 - (k_1 \cdot k_2))] \nonumber
     \\ \nonumber
     \\ \nonumber
    &= -2 \beta (1 + (1-\beta)f(\beta))
\end{align}
Where:

\begin{align}
    f(\beta) = \begin{cases} \mathrm{arcsin^2(\beta^{-1/2})} \quad \mathrm{for \beta \ge 1 } 
    \\
    \frac{-1}{4} \left[ln\frac{1 + \sqrt{1- \beta}}{1- \sqrt{1- \beta}} - i \pi \right] \mathrm{for \beta < 1} \end{cases}
\end{align}

The full matrix element for the W-boson loop diagrams in \ref{Diagrams} (a) and (b) are given by:

\begin{align}
    i \mathcal{M}_w = \int &\frac{d^dl}{(2 \pi)^d} (i \mathcal{M}_1 g^{\mu \nu} + i \mathcal{M}_2 l^{\mu} l^{\nu} + i \mathcal{M}_3 l^\mu k_1^\nu + i \mathcal{M}_3 l^\mu k_1^\nu + i \mathcal{M}_4 k_2^\mu l^\nu + i \mathcal{M}_5 k_1^\mu k_1^\nu) \epsilon_{*\mu}(k_1) \epsilon_\nu (k_2) \nonumber
    \\
    & \cdot \frac{1}{t^2-m_H^2 +im_H\Gamma_H} \cdot \left[\bar{u}(p_2) \frac{-ig_w}{2m_w} m_\chi u(p_1) \right]  
\end{align}

Where $\mathcal{M}_n$ are:

\begin{align}
i \mathcal{M}_1 &= -\frac{2e^2 g_w}{m_w^3 (l^2 - m_w^2)[(l - k_1)^2 - m_w^2][(l - k_1 - k_2)^2 - m_w^2]} \notag \\
& \quad \times \left\{ 6(l \cdot k_1)^3 - 2(l \cdot k_1)(l \cdot k_2)^2 + 2(l^2 - 3m_w^2)(l \cdot k_1)(l \cdot k_2) \right. \notag \\
& \quad - 3(l^2 - m_w^2)(l \cdot k_1)^2 + (l^2 - m_w^2)(l \cdot k_2)^2 + \left[ (l^2 - m_w^2)^2 + 2(1 - d)m_w^4 \right](l \cdot k_1) \notag \\
& \quad \left. - (l^2 - m_w^2)(l \cdot k_1) + m_w^2 \left[ ((d - 1)m_w^2 + m_H^2)(l^2 - m_w^2) + 4m_w^2 m_H^2 \right] \right\}  
\end{align}

\begin{equation}
i \mathcal{M}_2 = \frac{4e^2 g_w}{m_w (l^2 - m_w^2)[(l - k_1)^2 - m_w^2][(l - k_1 - k_2)^2 - m_w^2]} \left[ m_H^2 + 2(n - 1)m_w^2 \right]  
\end{equation}

\begin{align}
i \mathcal{M}_3 &= \frac{e^2 g_w}{m_w^3 (l^2 - m_w^2)[(l - k_1)^2 - m_w^2][(l - k_1 - k_2)^2 - m_w^2]} \notag \\
& \quad  \times \left\{ [(l^2)^2 - 3l^2 - 3m_w^2)(l \cdot k_1) - (l^2 + 7m_w^2)(l \cdot k_2) - 5l^2 m_w^2 \right. \notag \\
& \quad \left. + 2(l \cdot k_1)(l \cdot k_2) + 2(l \cdot k_1)^2 - 4(2d - 3)m_w^4 \right\}  
\end{align}

\begin{align}
i \mathcal{M}_4 &= -\frac{e^2 g_w}{m_w^3 (l^2 - m_w^2)[(l - k_1)^2 - m_w^2][(l - k_1 - k_2)^2 - m_w^2]} \notag \\
& \quad \times \left\{ (l^2 - m_w^2)(l^2 - 4m_w^2) - (3l^2 - 17m_w^2)(l \cdot k_1) - (l^2 - m_w^2)(l \cdot k_2) \right. \notag \\
& \quad \left. + 2(l \cdot k_1)(l \cdot k_2) + 2(l \cdot k_1)^2 \right\}  
\end{align}

\begin{equation}
i \mathcal{M}_5 = \frac{4e^2 g_w}{m_w (l^2 - m_w^2)[(l - k_1)^2 - m_w^2][(l - k_1 - k_2)^2 - m_w^2]} \left[ \frac{l^2 + 3m_w^2}{2} \right] 
\end{equation}

We rewrite $\mathcal{M}_1$, $\mathcal{M}_3$, and $\mathcal{M}_4$ as:

\begin{align}
i \mathcal{M}_1 &= \frac{2e^2 g_w}{m_w^3} \left[ \frac{2l \cdot (k_1 - k_2) + m_H^2}{4(l^2 - m_w^2)} - \frac{m_w^2}{(l - k_1)^2 - m_w^2} - \frac{2l \cdot (k_1 - k_2) - m_H^2 - 4m_w^2}{4[(l - k_1 - k_2)^2 - m_w^2]} \right. \notag \\
& \quad - \frac{4(m_H^2 + 2m_w^2)(l \cdot k_2) - 4(1 - d)m_w^4 - m_H^4}{4(l^2 - m_w^2)[(l - k_1 - k_2)^2 - m_w^2]} \notag \\
& \quad \left. - \frac{4m_H^2 m_w^4}{(l^2 - m_w^2)[(l - k_1)^2 - m_w^2][(l - k_1 - k_2)^2 - m_w^2]} \right] 
\end{align}

\begin{align}
i \mathcal{M}_3 &= \frac{e^2 g_w}{m_w^3} \left[ \frac{1}{(l^2 - m_w^2)[(l - k_1)^2 - m_w^2]} + \frac{2[(l - k_1 - k_2)^2 - m_w^2]}{(l^2 - m_w^2)[(l - k_1)^2 - m_w^2]} \right. \notag \\
& \quad + \frac{4m_w^2}{(l^2 - m_w^2)[(l - k_1)^2 - m_w^2]} + \frac{1}{(l^2 - m_w^2)[(l - k_1 - k_2)^2 - m_w^2]} \notag \\
& \quad \left. + \frac{(l^2 - m_w^2)}{[(l - k_1)^2 - m_w^2][(l - k_1 - k_2)^2 - m_w^2]} \right] 
\end{align}

\begin{align}
i \mathcal{M}_4 &= \frac{e^2 g_w}{m_w^3} \left[ \frac{2(l^2 - m_w^2)}{(l - k_1)^2 - m_w^2} - \frac{2[(l - k_1 - k_2)^2 - m_w^2]}{[(l - k_1)^2 - m_w^2]} \right. \notag \\
& \quad \left. + \frac{4m_w^2}{[(l - k_1)^2 - m_w^2][(l - k_1 - k_2)^2 - m_w^2]} \right]  
\end{align}

All of these components can be expanded using Passarino-Veltmann integrals, which can then be reduced to scalar integrals \cite{Marciano}. 

The final form of the matrix element is given by:
\begin{align}
    \mathcal{M}_w = \frac{e^2 g_w}{(4\pi)^2 m_w} I_w [(k_1 \cdot k_2) g^{\mu \nu} - k_2^\mu k_1^\nu] \epsilon_{*\mu}(k_1) e_\nu (k_2)
\end{align}

Where $\beta_w$ is:

\begin{align}
\beta_w = \frac{2{m_w^2}}{(k_2 \cdot k_1)} = -\frac{4 m_w^2}{t} 
\end{align}

And $I_w$ is given by:

\begin{align}
    I_w = 2+ 3\beta_w + 3(2\beta_w - \beta_w^2) f(\beta_w)
\end{align}

Finally we can now add the quark and W-boson loop contributions to get the full electroweak $\gamma \chi$ scattering matrix element:

\begin{align}
    i \mathcal{M} &= i \mathcal{M}_w + i \mathcal{M}_q \nonumber
    \\
    & = \frac{ie^2g_w}{(4\pi)^2 m_w} [(k_1 \cdot k_2)g^{\mu \nu} - k_2^\mu k_1^\nu] \epsilon^*_{\mu}(k_1)\epsilon_\nu (k_2) \frac{i}{t-m_H^2 + im_H \Gamma_H} \left[\bar{u}(p_2) \frac{-ig_w}{2m_w} m_\chi u(p_1) \right] (I_w + N_c Q_f^2 I_f) \nonumber
    \\
     \mathcal{M}^2&= \frac{e^4 g_w^4}{(4\pi)^4 m_w^4} \frac{3(k_1 \cdot k_2)^2}{(t-m_H^2)^2 + m_H^2 \Gamma_H^2} \frac{m_\chi^2 (2 m_\chi^2 - t/2)}{2} |I_w + N_c Q_f^2 I_f|^2 \nonumber
    \\
    &= \frac{e^4 g_w^4}{(4\pi)^4 m_w^4} \frac{3t^2}{8} \frac{m_\chi^2 (2 m_\chi^2 - t/2)}{(t-m_H^2)^2 + m_H^2 \Gamma_H^2} |I_w + N_c Q_f^2 I_f|^2
\end{align}
Where we have plugged in $p_1 \cdot p_2 = m^2_\chi - t/2$ and $k_2 \cdot k_1 = -t/2$. We multiply $\mathcal{M}^2$ with the phase-space and divide by the flux to obtain the differential cross section:
\begin{align}
   \frac{d\sigma}{d\Omega} &= \frac{\mathcal{M}^2} {64 \pi^2s} \nonumber
   \\
   &= \frac{\alpha^2 g_w^4}{(4\pi)^2 m_w^4} \frac{3t^2}{8} \frac{m_\chi^2 (2m_\chi^2 - \frac{t}{2})}{(t-m_H^2)^2 + m_H^2 \Gamma_H^2} \frac {\abs{I_w + N_c Q_f^2 I_f}^2} {64\pi^2 s}
\end{align}
Where $s$ is the Mandelstam variable. 

\subsubsection{Gravitationally Interacting DM}

In the following section will discuss how to obtain the graviton propagator and calculate $\gamma\chi$ scattering cross-section with graviton exchange. We begin with the Einstein-Hilbert action (S) for gravity, which is given by:

\begin{align}
    S = \frac{1}{16\pi G} \int d^4 x R \sqrt{-\mathrm{det}g_{\mu \nu}} 
\end{align}

where G is Newton's constant and R is the scalar curvature ($R \equiv g^{\mu \nu}R_{\mu \nu}$ where $R_{\mu \nu}$ is the Ricci tensor). It is a standard GR exercise to vary the action (or to consider what happens to S if we perturb the metric by an infinitesimal amount).  After performing these manipulations, we can recover the famous Einstein field equations:

\begin{align}
    R_{\mu \nu} - \frac{1}{2} g_{\mu \nu} R = -8 \pi G T_{\mu \nu}
\end{align}

where $T^{\mu \nu}$ is the stress-energy tensor, which encodes the matter fields of the universe. Let us assume the  weak-field approximation $g_{\mu \nu} = \eta_{\mu \nu}+\kappa h_{\mu \nu}$,where $\eta_{\mu \nu}$ is the flat Minkowski metric, $\kappa$ is a constant and $h_{\mu \nu}$ is the perturbation from the flat metric. With this assumption we may expand the action in terms of $h_{\mu \nu}$. This linearizes the effect of gravity. We incorporate $T^{\mu \nu}$ to include the graviton coupling to matter in our action:

\begin{align}
    S' = S - \int d^4 x \frac{1}{2} h_{\mu \nu} T^{\mu \nu}
\end{align}
    
Expanding the complete action to second order $O(h^2)$ and keeping careful track of the indices, we get the form:

\begin{align}
    S' = \int d^4 x \left( \frac{1}{32\pi G}I - \frac{1}{2}h_{\mu \nu} T^{\mu \nu}\right)
\end{align}

Where $I$ is the invariant quantity:

\begin{align}
    I \equiv \frac{1}{2} \partial_\lambda h^{\mu \nu} \partial^\lambda h_{\mu \nu} - \frac{1}{2}\partial_\lambda h^\mu_\mu \partial^\lambda h^\nu_\nu - \partial_\lambda h^{\lambda \nu} \partial^\mu h_{\mu \nu} + \partial^\nu h^\lambda_\lambda  \partial^\mu h_{\mu \nu}
\end{align}

If we then impose the harmonic gauge condition:

\begin{align}
    \partial_{\mu} h^\mu_\nu = \frac{1}{2} \partial_\nu h^\lambda_\lambda
\end{align}

which kills the second and last terms of $I$. In this harmonic gauge, the action becomes:

\begin{align}
    S' &= \int d^4 x \frac{1}{2} \left( \frac{1}{32 \pi G} \left( \partial_\lambda h^{\mu \nu} \partial^\lambda h_{\mu \nu} - \frac{1}{2}\partial_\lambda h^\mu_\mu \partial^\lambda h^\nu_\nu\right) - h_{\mu \nu}T^{\mu \nu}\right) \nonumber
    \\ 
    & = \frac{1}{32 \pi G} \int d^4 x (h^{\mu \nu} K_{\mu \nu;\lambda \sigma}(-\partial^2)h^{\lambda \sigma} - h_{\mu \nu}T^{\mu \nu}).
\end{align}

Here $ K_{\mu \nu;\lambda \sigma} \equiv \frac{1}{2}(\eta_{\mu \lambda} \eta_{\nu \sigma} + \eta_{\mu \sigma} \eta_{\nu \lambda} - \eta_{\mu \nu} \eta_{\lambda \sigma})$ is the matrix we must invert to build the graviton propagator \cite{Zee:2010Nutshell}, in analogy to the EM propagator. We note that the graviton field $h_{\mu \nu}$ is coupled to the stress energy tensor, $T^{\mu \nu}$, as the stress-energy tensor encodes all the energy and matter in the universe, this reflects everything is subject to the graviton field and the effect of gravity, as expected from GR. From the convenient fact that $K= K^{-1}$, we find the propagator:

\begin{align}
    D_{\mu \nu ; \lambda \sigma} (k) = \frac{1}{2} \frac{\eta_{\mu \lambda} \eta_{\nu \sigma} + \eta_{\mu \sigma}\eta_{\nu \lambda} - \eta_{\mu \nu} \eta{\lambda \sigma}}{k^2 + i \varepsilon}
\end{align}

where $k$ is the momentum transferred by the graviton exchange and $\varepsilon$ is an infinitesimal quantity.
We begin with the Lagrangian for an electromagnetic field coupled to the graviton field:

\begin{align}
    \mathcal{L} = \sqrt{-\mathrm{det}g_{\mu \nu}} \left(\frac{-1}{4}g^{\mu \nu}g^{\alpha \beta} F_{\mu \alpha } F_{\nu \beta} \right) 
\end{align}
Here, $F_{\mu \nu}$ is the EM tensor:

\begin{align}
    F^{\mu \nu} = \partial^\mu A^\nu - \partial^\nu A^\mu
\end{align}

\begin{figure}[!h]
\begin{center}
\includegraphics[width=0.5\textwidth,angle=0]{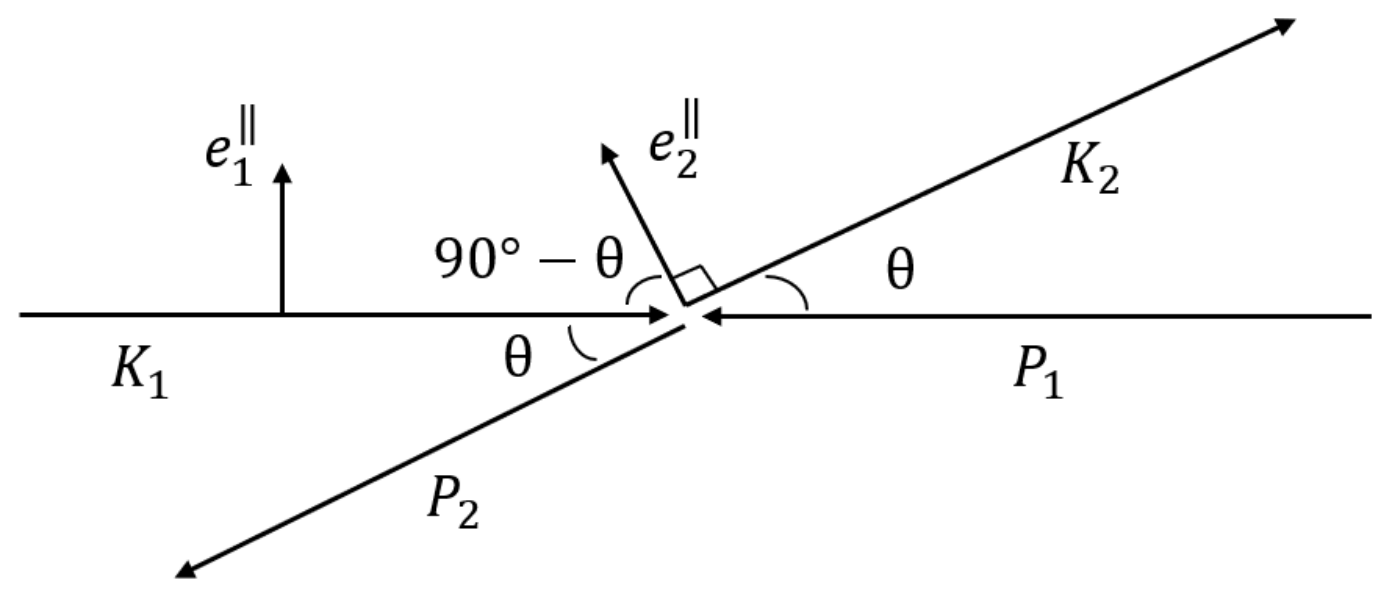}
\includegraphics[width=0.5\textwidth,angle=0]{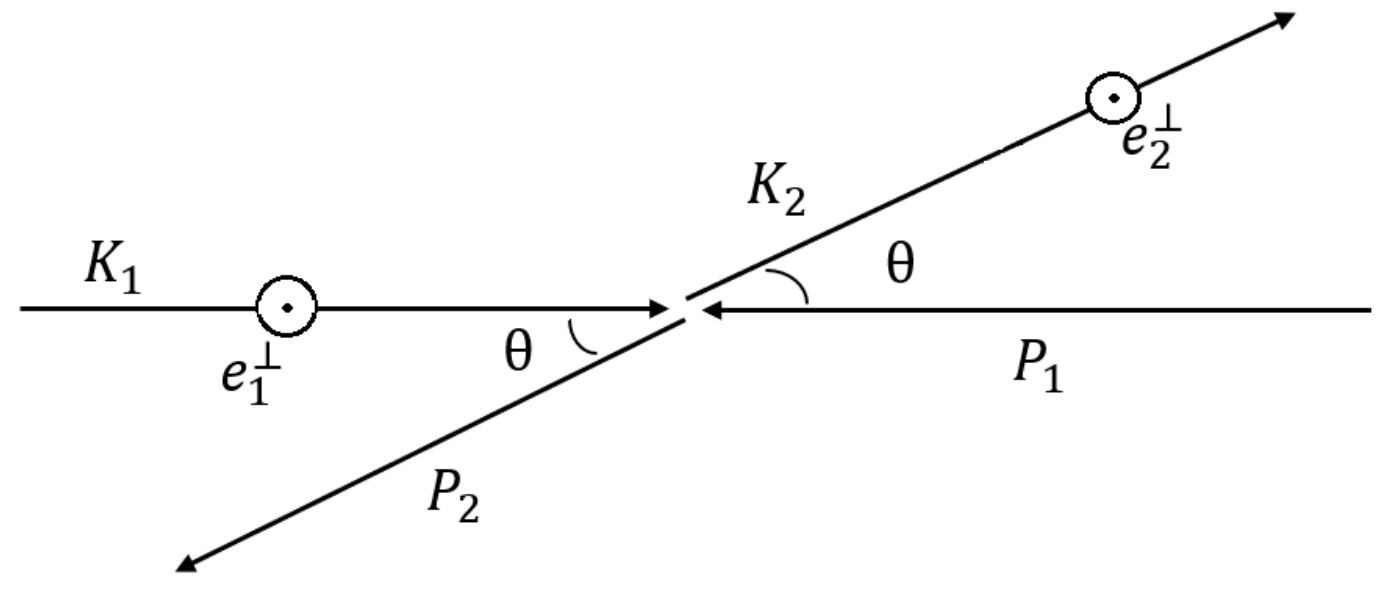}
\end{center}
\caption{Rotation of polarization basis vector in the scattering plane (top) and orthogonal to the scattering plane (bottom).}
\label{PolarizationVectors}
\end{figure}

In our weak-field approximation, we have $g_{\mu \nu} = \eta_{\mu \nu}+\kappa h_{\mu \nu}$. This linear perturbation represents the effect of gravity as seen by the metric. When we plug in the weak field approximation in the Lagrangian we get:

\begin{align}
     \sqrt{-\mathrm{det}g_{\mu \nu}} \left(\frac{-1}{4}g^{\mu \nu}g^{\alpha \beta} F_{\mu \alpha } F_{\nu \beta} \right) &\approx \sqrt{-\mathrm{det}g_{\mu \nu}} \left(\frac{-1}{4} \right) (\eta^{\mu \nu} \eta^{\alpha \beta}- \kappa h^{\mu \nu} \eta^{\alpha \beta} - \kappa h^{\alpha \beta} \eta^{\mu \nu}) F_{\mu \alpha}F_{\nu \beta} \nonumber
     \\
     & = \sqrt{-\mathrm{det}g_{\mu \nu}}\left(\frac{-1}{4} \right) (\eta^{\mu \nu} \eta^{\alpha \beta}F_{\alpha \mu } F_{\beta \nu} + \kappa h^\mu_\nu \eta^{\nu \nu} \eta^{\alpha \beta}F_{\mu \alpha} F_{\beta \nu} + \kappa h^\alpha_\beta \eta^{\beta \beta} \eta^{\mu \nu} F_{\mu \alpha} F_{\beta \nu}) \nonumber
     \\
     & = \sqrt{-\mathrm{det}g_{\mu \nu}}\left(\frac{-1}{4} \right)(F^{\mu \nu} F_{\mu \nu} + 2 \kappa h^\mu_\nu F_{\mu \alpha}F^{\alpha \nu}) \nonumber
     \\
     & = \left( 1 + \frac{\kappa}{2}\eta_{\alpha \beta} h^{\alpha \beta}\right) \left[ \frac{-1}{4} (F^{\mu \nu}F_{\mu \nu} + 2 \kappa h^\mu_\nu F_{\mu \rho}F^{\rho \nu})\right]
\end{align}
Where to obtain the last line, we use $ \sqrt{-\mathrm{det}g_{\mu \nu}} \approx 1 +\frac{\kappa}{2} \eta_{\alpha \beta} h^{\alpha \beta}$. We distribute and collect the terms for the interaction Lagrangian, keeping only terms first order in $\kappa$, and we obtain 6 terms:

\begin{align}
    \mathcal{L}_{int} = -\frac{\kappa}{4} \eta_{\alpha \beta} &h^{\alpha \beta} \partial_\mu A_\nu \partial^\mu A^\nu + \frac{\kappa}{4} \eta_{\alpha \beta} h^{\alpha \beta} \partial_\mu A_\nu \partial^\nu A^\mu - \frac{\kappa}{2} h^{\mu \nu} \partial_\mu A_\rho \partial^\rho A_\nu  \nonumber
    \\
    &+ \frac{\kappa}{2} h^{\mu \nu} \partial_\rho A_\mu \partial^\rho A_\nu + \frac{\kappa}{2} h^{\mu \nu} \partial_\mu A_\rho \partial_\nu A^\rho - \frac{\kappa}{2} h^{\mu \nu } \partial_\rho A_\mu \partial_\nu A^\rho
\end{align}
We use the Feynman rules derived for a spin-2 particle in the massless limit to calculate the photon-photon-graviton vertex \cite{Prinz2021, Bern2002}.

\begin{align}
    V_{k_1,k_2} = \frac{i\kappa}{2}[k_1 \cdot k_2(\eta^{\rho \alpha} \eta^{\beta \sigma} - \eta^{\rho \beta} \eta^{\alpha \sigma} - \eta^{\rho \sigma} \eta^{\alpha \beta}) + \eta^{\rho \sigma}k_1^{\beta}k_2^\alpha + \eta^{\alpha \beta}k_1^\sigma k_2^\rho - (\eta^{\alpha \rho} k_2^\sigma k_1^\beta + \eta^{\beta \rho}k_1^\sigma k_2^\alpha - \eta^{\beta \sigma}k_1^\rho k_2^\alpha)]
\end{align}
We now consider the diagram in Figure \ref{Diagrams} (c), and choose basis vectors for the photon polarization. As is typical \cite{Peet1970}, we choose one vector,
$e_{\parallel}$ within the plane spanned by $k_1, k_2$, and the other vector, $e_{\perp}$, we choose as orthogonal to this scattering plane, as shown in Figure \ref{PolarizationVectors}. Following the conventions in \cite{Boccaletti1969}, we write:

\begin{align}
e_1^{\perp} &= (0,1,0)
\\
e_2^{\perp} &= (0,1,0)
\\
e_1^{\parallel} &= (1,0,0)
\\
e_2^{\parallel} &= (-cos\theta , 0, sin\theta)
\end{align}

We assume the gauge condition $e^0=0$. And then, we can write the matrix element as:

\begin{align} \label{m}
    \bra{k_2,p_2}M\ket{k_1,p_1}&=\frac{\lambda^2\delta^4(k_2+p_2-k_1-p_1)}{4(2\pi)^2\sqrt{k_1^0k_2^0p_1^0p_2^0}}\nonumber \frac{4}{-2k_1k_2}\cdot
    \{(e_{(\rho}^2k_\alpha^2-e_\alpha^2 k_\rho^2)(e_\sigma^1k^\alpha_1-e_1^\alpha k_{\sigma)}^1)\nonumber\\
    &-\frac{1}{4}\delta_{\rho\sigma}(e^2_\alpha k^2_\beta-e^2_\beta k^2_\alpha)(e^\alpha_1k^\beta_1-e^\beta_1k^\alpha_1)\} \cdot\left[p^{(\rho}_1p^{\sigma)}_2-\frac{1}{2}m^2\delta^{\rho\sigma}\right]
\end{align}

 where Greek indices represent components of 4-vectors and the numbers $1,2$ correspond to initial and final particles respectively. The quantity $\lambda=\sqrt{8\pi G}$ governs the strength of the gravitational interaction and $\delta^{\rho\sigma}$ is the usual Kronecker delta and $A_{(\rho}B_{\sigma)}$ denotes symmetrization with respect to the indices $\rho$ and $\sigma$. The symmetrization of the expression in the curly brackets in \ref{m} is less trivial:

 \begin{align} \label{pol}
      (e_{(\rho}^2k_\alpha^2-e_\alpha^2 k_\rho^2)(e_\sigma^1k^\alpha_1-e_1^\alpha k_{\sigma)}^1) = \frac{1}{2} [(e_{\rho}^2k_\alpha^2-e_\alpha^2 k_\rho^2)(e_\sigma^1k^\alpha_1-e_1^\alpha k_{\sigma}^1) + (e_\sigma^2k^\alpha_2-e_2^\alpha k_{\sigma}^2) (e_{\rho}^1k_\alpha^1-e_\alpha^1 k_\rho^1)]
 \end{align}

 Substituting this back into \ref{pol}, we distribute and simplify by contracting the 4-vectors having equivalent indices. This contraction produces several occurrences of the scalar products $p_1 \cdot e_1, p_2 \cdot e_2, k_1 \cdot e_1$ and $k_2 \cdot e_2$, which are equal to zero in CoM, since the polarization of light is transverse to the direction of motion. Including only non-zero terms, and grouping like terms, we obtain the full expression:
 
\begin{align} \label{m2}
  \bra{k_2,p_2}M\ket{k_1,p_1}&=\frac{\lambda^2\delta^4(k_2+p_2-k_1-p_1)}{4(2\pi)^2\sqrt{E_{k_1} E_{k_2} E_{p_1}E_{p_2}}}  \frac{1}{k_1 \cdot k_2} ((p_1\cdot e_2)(p_2 \cdot e_1)(k_1 \cdot k_2) - (p_1\cdot e_2)(p_2\cdot k_1)(k_2\cdot e_1) \nonumber
  \\
  &-(p_1\cdot k_2)(p_2 \cdot e_1)(k_1 \cdot e_2) + (p_1 \cdot k_2)(p_2 \cdot k_1)(e_1\cdot e_2)) - ((p_1p_2)(k_1 \cdot e_2)(k_2 \cdot e_1)\nonumber
  \\
  &-(p_1 \cdot p_2)(k_1 \cdot k_2)(e_1 \cdot e_2) )  
\end{align}
Where $E_{k1,k2,p1,p2}$ is the energy of the concerned particle indicated in the subscript. Since we are working in the CoM frame we have:

\begin{align}
    E_{k_1} &=  E_{k_2} = E_k
    \\
    E_{p_1} &= E_{p_2} = E_p
\end{align}
By plugging our kinematics into \ref{m2}, we obtain for the sum in curly brackets (for some specified initial and final polarization states).

\begin{align}
e_1^{\perp} \rightarrow e_2^{\perp} &= E^2_{k}(E_k + E_p)^2 
\\
e_1^{\perp}  \rightarrow e_2^{\parallel} &= 0
\\
e_1^{\parallel}  \rightarrow e_2^{\parallel} &=  E_k^2(E_k^2 + E_p^2)^2  \mathrm{cos}\theta
\\
e_1^{\parallel} \rightarrow e_2^{\perp} &= 0
\end{align}
Thus our non-zero, polarization-dependent matrix elements become:
\begin{align}
 \bra{k_2,p_2}M_{\parallel \parallel}\ket{k_1,p_1}&= \frac{-\lambda^2 \delta^4(k_2 + p_2 -k_1 - p_1) (E_k+E_p)^2}{4(2\pi)^2 E_p}\frac{\mathrm{cos}\theta}{1- \mathrm{cos}\theta} 
 \\
 \bra{k_2,p_2}M_{\perp \perp}\ket{k_1,p_1}&= \frac{-\lambda^2 \delta^4(k_2 + p_2 -k_1 - p_1) (E_k+ E_p)^2}{4(2\pi)^2 E_p}\frac{1}{1- \mathrm{cos}\theta} 
\end{align}

We can obtain the polarization-dependent differential cross sections as:
\begin{equation}
    \begin{split}
        \frac{d\sigma_{\parallel\parallel}}{d\Omega}=\frac{\lambda^4(E_k+E_p)^4}{16(2\pi)^2E_p^2(1-\text{cos}\theta)^2}\\
        \frac{d\sigma_{\perp\perp}}{d\Omega}=\frac{\lambda^4(E_k+E_p)^4\text{cos}^2\theta}{16(2\pi)^2E_p^2(1-\text{cos}\theta)^2}
    \end{split}
\end{equation}

\end{document}